\documentclass[aps,prl,reprint,superscriptaddress]{revtex4-1}
\usepackage{graphicx}
\usepackage[utf8]{inputenc}

\makeatletter
\def\blfootnote{\xdef\@thefnmark{}
\@footnotetext}
\makeatother
\begin{document}

%\preprint{}

\title{Impact of $^{16}$O($\gamma$,$\alpha$)$^{12}$C measurements on the $^{12}$C($\alpha,\gamma$)$^{16}$O astrophysical reaction rate}


\author{R. J. Holt}
\affiliation{Physics Division, Argonne National Laboratory, Argonne, Illinois 60439}
\affiliation{Kellogg Radiation Laboratory, California Institute of Technology, Pasadena, California 91125}
\email{rholt@caltech.edu}
%\thanks{This work is supported by the U.S. National Science Foundation under grant 1506459 and by the U.S. Department of Energy (DOE), Office of Science, Office of Nuclear Physics, under contract No. DE-AC02-06CH11357.}
\author{B. W. Filippone}
\affiliation{Kellogg Radiation Laboratory, California Institute of Technology, Pasadena, California 91125}
\email{bradf@caltech.edu}
\author{Steven C. Pieper\,\textdagger}
\blfootnote{\textdagger\,Deceased}

\affiliation{Physics Division, Argonne National Laboratory, Argonne, Illinois 60439}
%\email[]{spieper@anl.gov}


\date{\today}

\begin{abstract}
The $^{12}$C($\alpha,\gamma$)$^{16}$O reaction, an important component of stellar helium burning, plays a key role in nuclear astrophysics. It has direct impact on the evolution and final state of massive stars, while also influencing the elemental abundances resulting from nucleosynthesis in such stars.  Providing a reliable estimate for the energy dependence of this reaction at stellar helium burning temperatures has been a major goal for the field. In this work, we study the role of potential new measurements of the inverse reaction, $^{16}$O($\gamma,\alpha$)$^{12}$C, in reducing the overall uncertainty. A multilevel $R$-matrix analysis is used to make extrapolations of the astrophysical $S$ factor for this reaction to the stellar energy of 300 keV. The statistical precision of the $S$-factor extrapolation is determined by performing multiple fits to existing $E1$ and $E2$ ground state capture data, including the impact of possible future measurements of the $^{16}$O($\gamma,\alpha$)$^{12}$C reaction. In particular, we consider a proposed Jefferson Laboratory (JLab) experiment that will make use of a high-intensity low-energy bremsstrahlung beam that impinges on an oxygen-rich single-fluid bubble chamber in order to measure the total cross section for the inverse reaction. The importance of low energy data as well as high precision data is investigated.
\end{abstract}

%\pacs{}
%\keywords{}

\maketitle

\section{Introduction}
The $^{12}$C($\alpha,\gamma$)$^{16}$O reaction is believed to be one of the most important reactions in nuclear astrophysics\cite{Fowler:1984zz,Woosley:2003nki}. A recent review\cite{deBoer:2017ldl} highlights the key role played by this reaction in both the evolution of and nucleo-synthetic yields from massive stars. The purpose of this study is to explore the role that forthcoming measurements of the inverse reaction - $^{16}$O($\gamma,\alpha$)$^{12}$C  (OSGA) - could have on reducing the overall uncertainty in the cross section for the $^{12}$C($\alpha,\gamma$)$^{16}$O reaction at helium burning temperatures. To do this we perform fits to the existing data using the $R$-matrix approach\cite{Lane:1948zh} and study the impact of including new data on the inverse reaction. This is achieved by starting with a reasonable $R$-matrix fit that can be used as a basis for comparison to fits with and without projected $^{16}$O($\gamma,\alpha$)$^{12}$C data. For the inverse capture data we start with a proposed JLab experiment\cite{suleiman:2014aa} in order to assess the possible role of new measurements in reducing the overall uncertainty in the cross section\cite{Holt:2018cgp}.   A detailed $R$-matrix analysis of this reaction and and excellent review of the subject is given in Ref.\cite{deBoer:2017ldl}.  

In the present work, we employed the $R$-matrix approach to calculate the total cross section, $\sigma(E)$, for alpha-capture to the ground state.  Considering only ground state capture is sufficient for this study since the capture to excited states is believed\cite{deBoer:2017ldl} to contribute only about 5$\%$ to the total capture rate at 300 keV.  The cross section is then used to calculate the astrophysical $S$ factor given by

\begin{equation}
S(E) = \sigma(E)Ee^{2\pi \eta}
\end{equation}
where $E$ is the energy in the center of mass, $\eta$ is the Sommerfeld parameter, $\sqrt{\frac{\mu}{2E}}Z_1Z_2\frac{e^2}{\hbar}$, and $\mu$ is the reduced mass of the carbon ion and alpha particle.  Measurements of the $S$ factor as a function of energy are often reported in the literature.  For the $^{12}$C($\alpha,\gamma$)$^{16}$O reaction, the value of $S$ at $E=300\ keV$ is typically quoted as the most probable energy for stellar helium burning.     Of course, the cross section is so small at 300 keV that it cannot be directly measured.  Thus, extrapolations to 300 keV must be performed to study the impact of data on the extrapolation.  Of course, efforts aimed at improving the data and extrapolation are underway\cite{suleiman:2014aa,Gai:2018sip,Balabanski:2017qth,Costantini:2009wn,Robertson:2016llv,Liu:2017zvh,Bemmerer:2018zsh,xu:2007,Friscic:2019eow} at a number of laboratories worldwide.   The new inverse reaction (OSGA) experiments\cite{suleiman:2014aa,xu:2007,Gai:2018sip,Balabanski:2017qth,Friscic:2019eow} bring a different set of systematic errors than previous experiments and thus provide an additional check on systematics.

\section{$R$-matrix approach}

The collision  matrix for the OSGA reaction will be given in terms of the Hamiltonian $H^{\mathcal{L}}$ which electromagnetically couples the photon of multipolarity $\mathcal{L}$ to the nucleus.  We introduce the wave function $\Psi_{E(J)}$ that describes the alpha-$^{12}$C system in total spin state $J$ and an initial state wave function $\psi_{i(J_i)}$ which describes the nucleus ($^{16}$O) in its ground state.  Then the collision matrix is given by

\begin{equation}
 U_{\gamma\mathcal{L} f,c}^{(J)}=\left[\frac{8\pi(\mathcal{L}+1)}{\mathcal{L}\hbar}\right]
^\frac{1}{2}\frac{k_\gamma^{\mathcal{L}+\frac{1}{2}}}
{(2\mathcal{L}+1)!!}
\frac{<\Psi_{E(J)|| H^{\mathcal{L}}||}\psi_{i(J_i)}>}{(2J + 1)^{\frac{1}{2}}}
\end{equation}
where $k_\gamma=E_\gamma/\hbar c$ is the photon wave number and the subscript c refers to the final $\alpha-^{12}$C channel with quantum numbers $slJ$.  Here $s$ is the channel spin (zero in this case), $l$ is the orbital angular momentum, and $J=l+s$ is the total angular momentum.  In principle, we would perform the radial integration in Eq. 2 from the origin to the channel radius (internal piece) and from the channel radius to infinity (external piece).
According to the R-matrix theory\cite{Lane:1948zh} inside the channel radius $a$, the final state wave function, $\Psi_{E(J)}$, can be expanded in terms of a complete set of states, $X_{\lambda(J)}$

\begin{equation}
\Psi_{E(J)} = i\hbar^{1/2}e^{-i\phi_c}\Sigma_{\lambda\mu}A_{\lambda\mu}\Gamma_{\mu c}^{1/2}X_{\lambda(J)}  
\end{equation}
where $\phi_c$ is a Coulomb phase shift, $\Gamma_{\mu c}$ is the width of level $\mu$ in channel $c$, and $A_{\lambda\mu}$ is the matrix that relates the internal wave function and the observed resonances.
Here 
\begin{equation}
\left(A^{-1}\right)_{\lambda\mu}=\left(E_\lambda - E\right)\delta_{\lambda\mu} - \xi_{\lambda\mu}  
\end{equation}
where $E_\lambda$ is a level energy, $\delta_{\lambda\mu}$ is the Kronecker $\delta$ and $\xi$ is given in terms of the Coulomb shift factor, $S_c$, the boundary condition constant, $b_c$, and the Coulomb penetration factor, $P_c$ 

\begin{equation}
\xi_{\lambda\mu} = \Sigma_c [(S_c - b_c) + iP_c]\gamma_{\lambda c}\gamma_{\mu c} 
\end{equation}
where here c refers to essentially the $\alpha$ channel in this case and the $\gamma_{\lambda c}$  are the $\alpha$ reduced width amplitudes.  The $\alpha$ channel is the only open channel and closed channels are neglected.

The internal part of the collision matrix for radiative capture to the ground state is given by

\begin{equation}
  U_{\gamma \alpha}^{lJ\mathcal{L}} = i e^{-i\phi_l} \Sigma_{\lambda\mu} A_{\lambda\mu}\Gamma_{\lambda\alpha l J}^{1/2}\Gamma_{\mu\gamma l J}^{1/2} 
%  + \left(\frac{8\pi(\mathcal{L} + 1)}{(2J+1)\mathcal{L}\hbar v_\alpha}\right)^{1/2} \frac{k_\gamma^{\mathcal{L}+1/2}}{(2\mathcal{L} + 1)!!}
%<\Psi_{f(J_f)}||H^\mathcal{L} ||\psi_i>  
\end{equation}
where $\phi_l$ is the Coulomb phase shift for orbital angular momentum $l$, $\Gamma_{\lambda\alpha lJ}$ and $\Gamma_{\mu\gamma lJ}$ are the formal ground state $\alpha$ and radiative widths, respectively.
For a given level, the observed width can be related\cite{Lane:1948zh} to the reduced width by
\begin{equation}
\Gamma_{\lambda \alpha l J}  = \frac{2 P_l\gamma^2_{\lambda \alpha l J}}{1 + \gamma^2_{\lambda \alpha l J}\left(\frac{dS_l}{dE}\right)}
\end{equation}
while the reduced widths for the bound states are given by
\begin{equation}
\gamma_{1 \alpha l b} ^2 = \frac{\gamma^2_{1 \alpha l}}{1 + \gamma^2_{1 \alpha l}\left(\frac{dS_{bl}}{dE}\right)}
\end{equation}
where $S_{bl}$ is the bound state shift factor for orbital angular momentum $l$.
For the photon radiative width, we have
\begin{equation}
\Gamma_{\lambda \gamma l J}  = P_{\gamma \lambda}\Gamma_{\lambda \gamma lJ\circ}\left[1 + \gamma^2_{\lambda \alpha l J}\left(\frac{dS_l}{dE}\right)\right]
\end{equation}
where $\Gamma_{\lambda \gamma lJ\circ}$ is the observed radiative width and
\begin{equation}
P_{\gamma \lambda} \equiv \left[\frac{E + Q}{E_{r\lambda} + Q}\right]^{(2\mathcal{L} + 1)/2}
\end{equation}
where $Q$ is the $Q$-value for the reaction and $E_{r\lambda}$ are the physical resonance energies as given in the equation $E_{r\lambda} = E_\lambda + (b_c - S_c)\gamma_{\lambda\alpha}^2$.

We then calculated the $E\mathcal{L}$ ground state radiative cross section\cite{Carr:1971rfo} for the $^{12}$C($\alpha$,$\gamma$)$^{16}$O reaction from the collision matrix for spin-zero nuclei:
\begin{equation}
\sigma_{E\mathcal{L}}(E) = \frac{(2\mathcal{L} + 1)\pi}{k_\alpha^2}| U_{\gamma \alpha}^{lJ\mathcal{L}}|^2
\end{equation}

%For a given level, the observed width can be related to the reduced width by
%\begin{equation}
%\Gamma^o_{\lambda c}  = \frac{2 P_c\gamma^2_{\lambda c}}{1 + \gamma^2_{\lambda c}\left(\frac{dS_c}{dE}\right)_{E=E_R}}
%\end{equation}
%For the observed photon radiative width, this expression becomes
%\begin{equation}
%\Gamma^o_{\lambda \gamma}  = \frac{\Gamma_{\lambda \gamma}}{1 + \gamma^2_{\lambda c}\left(\frac{dS_c}{dE}\right)_{E=E_R}}
%\end{equation}
%and the photon penetrability is given by $k_\gamma^{2\mathcal{L} + 1}$.

We only considered ground state transitions and statistical errors in this study.  We initially chose a channel radius of 5.43 fm to be consistent with a previous analysis\cite{deBoer:2017ldl}, but later consider a larger channel radius to be consistent with other analyses\cite{Kirsebom:2018amt,Shen:2018eis}.  We employed five $E1$ resonance levels and four $E2$ resonance levels in the internal part of the the R-matrix analysis as shown in Table~ \ref{one}.  This analysis is similar to that of refs.{\cite{Azuma:2010zz} and {\cite{Holt:1978zz}, and the details comport with the results of Lane and Thomas\cite{Lane:1948zh}.   In order to speed up computations, we turned off the external part for this study.  This external contribution is most sensitive to the $E2$ part of the cross section since the $E1$ external part is greatly reduced by isospin symmetry.  In fact, the external E1 part would vanish under perfect isospin conservation.  We performed the fit for data less than 3 MeV, where the external part is small.  As a check, we turned on the external piece for several fits, but it did not significantly change the results.

\begin{table}[!htp] %add [H] placement to break table across pages
\caption{\label{one}Parameters used in the present simultaneous fits to original data for $E1$ and $E2$, and a channel radius of 5.43 fm.  These parameters were used to generate the curves in Fig.~\ref{fig1}.  The $E_\lambda$ are eigenenergies not physical resonance energies.  The widths for resonances above threshold are the observable widths $\Gamma_{\lambda\alpha}$.  The widths for the bound states are reduced widths $\gamma^2_{1\alpha b}$.  The minus signs in front of the widths indicate the signs of the reduced width amplitudes.  The values marked with an asterisk were allowed to vary in the fit, and are given for the ``all" fit in Table~\ref{two}.  All other parameters were fixed.}
\begin{ruledtabular}
\begin{tabular}{c c c c  c c c}
   &    &    $E1$   &    &         &   $E2$  & \\
$\lambda$ &  E$_\lambda$  & $\Gamma_{\lambda\alpha }/\gamma^2_{1 \alpha  b}$  & $\Gamma_{\lambda\gamma \circ}$  &E$_\lambda$ &  $\Gamma_{\lambda\alpha}/\gamma^2_{1\alpha  b}$  & $\Gamma_{\lambda\gamma \circ}$  \\
   &   (MeV) & (keV) & (eV) & (MeV) & (keV) & (eV)\\
\hline
1  &   -0.297 &  114.6$^*$            &  0.055          &   -0.482    &   105.0$^*$    &  0.097  \\
2  &    2.416  & 414.7$^*$            &  -0.0152$^*$    &    2.683    &   0.62            &   -0.0057\\
3  &    5.298  &  99.2                  &  5.6              &    4.407      &   83.0          &  -0.65  \\
4  &    5.835  &  -29.9                 &  42.0                &    6.092        &   -349          &  -1.21$^*$  \\
5  &   10.07     & 500                    &  0.604$^*$        &     -           &    -               &   -   \\
\end{tabular}
\end{ruledtabular}
 \end{table}

\section{Simultaneous fits and projections for $S_{E1}$, $S_{E2}$ and total $S$ }

We used a SIMPLEX fitter\cite{Nelder:1965zz} for the present work.  Our best $R$-matrix fit of the existing $E1$ and $E2$ $S$-factor data, shown in Fig.~\ref{fig1}, was taken as the most probable description of the $S$-factor data.  In order to explore the statistical variation in the S-factor extrapolations, we created $S$-factor pseudo-data by random variation according to a Gaussian probability distribution about the best fit $S$-factor values at the measured energies.  In the randomizations, we multiplied the individual pseudo-data uncertainties by the square root of the ratio of the original best fit values to the original measured uncertainties.  We further multiplied these uncertainties by the square root of the $E1$ and $E2$ reduced chi squares, the Birge factor\cite{birge}, for the $E1$ and $E2$ fits, respectively.  This procedure should give a conservative estimate for statistical uncertainties.  For the subtheshold states, we fixed the radiative widths of the subthreshold states at the measured values and varied the reduced alpha widths.  We allowed the reduced alpha and radiative width of the first $E1$ state above threshold to vary in the fit, while we allowed the radiative width of the fifth $E1$ state to vary.   We also allowed the radiative width of the fourth $E2$ R-matrix level to vary.
 The first $E2$ state above threshold is very narrow and we fixed the parameters of this level at those of ref.{\cite{deBoer:2017ldl}.  The radiative width of the third $E2$ resonance was treated separately.  We observed that using the value in ref.{\cite{deBoer:2017ldl} resulted in a cross section that was significantly smaller than the data of ref.\cite{Schurmann:2011zz}.  Rather, we made a fit to $E2$ data that included the data of ref.\cite{Schurmann:2011zz}.  We then fixed the third $E2$ radiative width at -0.65 eV found from the fit and used it in subsequent fits to the data below 3 MeV.  Indeed, we fixed all other parameters except the third $E2$ radiative width and those marked with an asterisk in Table~\ref{one} at the values of ref.{\cite{deBoer:2017ldl}.  The parameters allowed to vary are denoted by an asterisk in table~\ref{one}.

 Also, following ref. {\cite{deBoer:2017ldl}}, we performed the fits by maximizing L rather than minimizing $\chi^2$, where L is given\cite{sivia} by
\begin{equation}
L = \Sigma_i ln[(1-exp(-R_i/2))/R_i]
\end{equation}
and $R_i=(f(x_i)-d_i)^2/\sigma_i^2$ is the usual quantity used in $\chi^2$ minimizations.  Here $f(x_i)$ is the function to be fitted to data, $d_i$, with statistical error $\sigma_i$.  The L maximization has the feature that it reduces the impact of large error bar data on the fit and generally gives larger S-factor uncertainties in projected values of $S(300\ keV)$ than that of a $\chi^2$ minimization.   In this work $L_{tot}$ is maximized and defined by

\begin{equation}
L_{tot} = L_{E1} + L_{E2} + L_{OSGA}
\end{equation}
where $L_{E1(2)}$ is $L$ for $E1(2)$ data and $L_{OSGA}$ represents $L$ for the inverse reaction data or JLab data in this case.

The parameters of the bound levels are very important for the projection to 300 keV. The resonance energies were fixed, but the parameters, $E_\lambda$, depend on the reduced width of the levels. We allowed the reduced widths of the bound states to vary, so the $E_\lambda$ varies.  We chose the $R$-matrix boundary condition constants to cancel out this effect for the second levels so that $E_\lambda=E_{r\lambda}$ for these levels. For the third and higher levels, the reduced widths were not varied because alpha elastic scattering determined these widths and allowing them to vary did not make a significant difference. 
 We used the $S$-factor data sets given in refs. \cite{Dyer:1974pgc,Kremer:1988zz,Redder:1987xba,Ouellet:1992zz,Roters:1999zz,Gialanella:2001ayx,Kunz:2001zz,Assuncao:2006vy,Makii:2009zz,Plag:2012zz} and show the E1 and E2 ground state S factors in Fig.\ref{fig1}.

\begin{figure}[ht]
%\begin{center}
\includegraphics[width=3.4in]{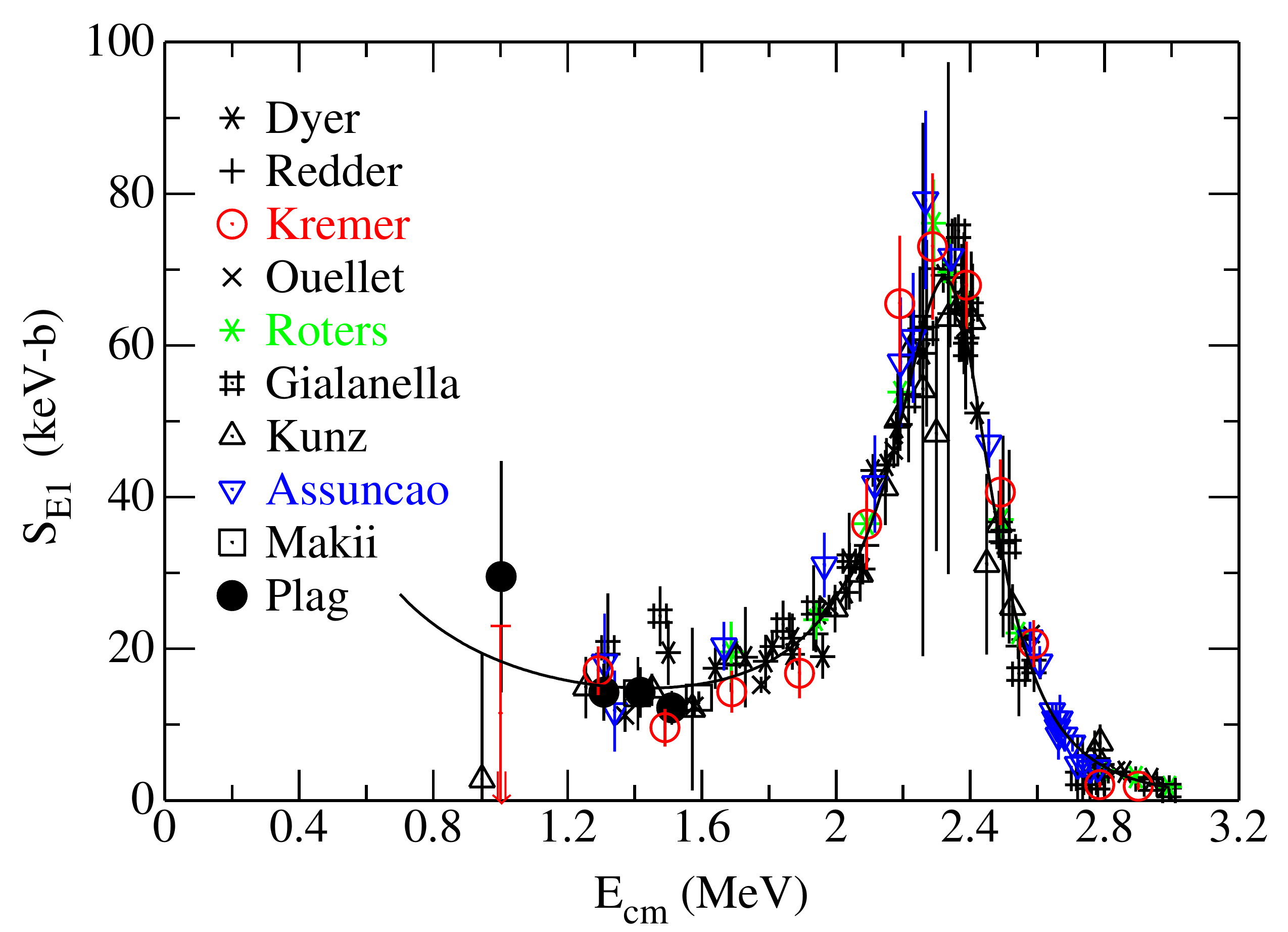}
\includegraphics[width=3.4in]{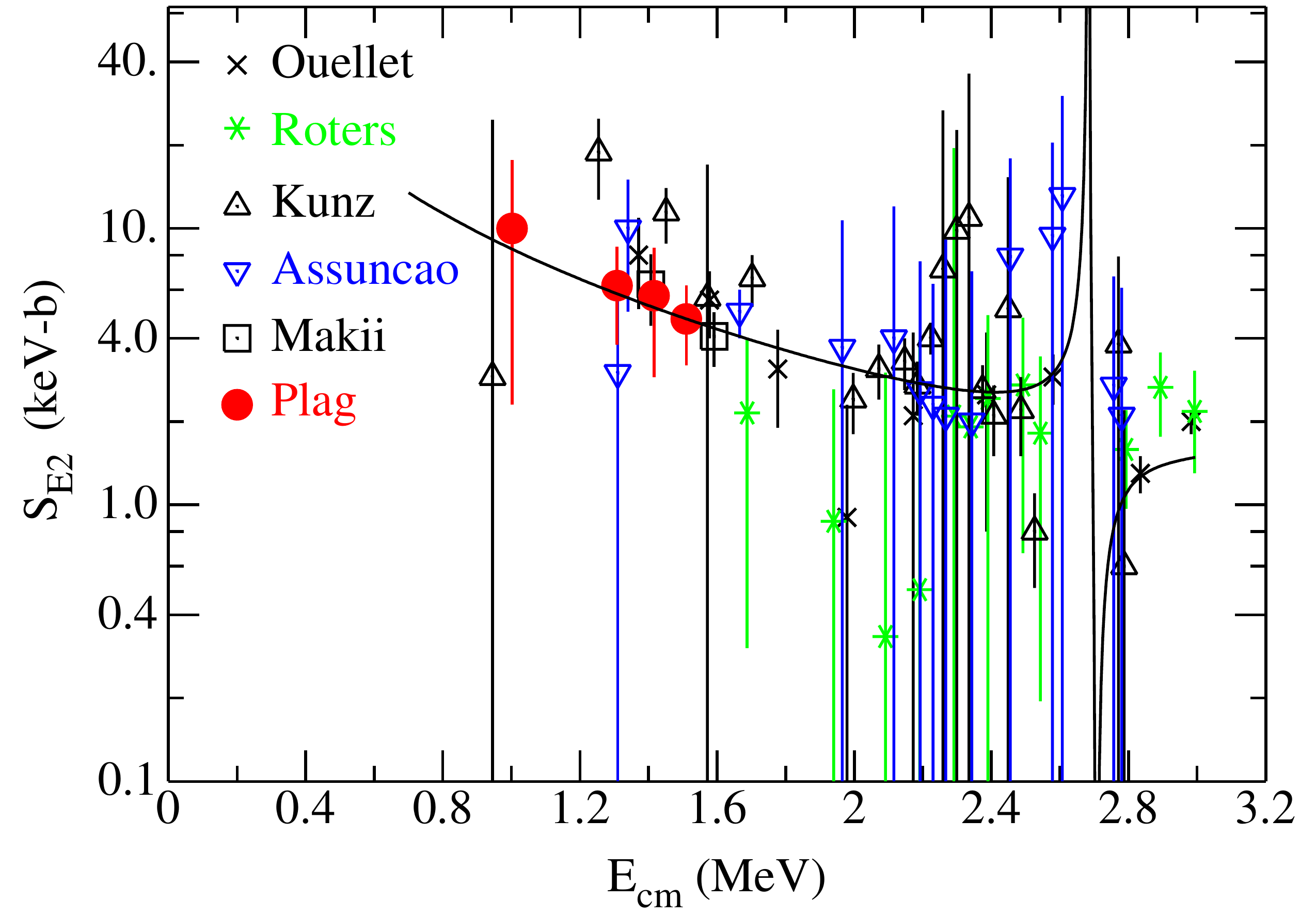}
\caption{The astrophysical $S$ factor for the $E1$ ($E2$) cross section as a function of center of mass energy is shown in the top (bottom) panel.  The solid curves represent the best fits and are based on the parameters in Table~\ref{one}, while the data are taken from the refs.~\cite{Dyer:1974pgc,Kremer:1988zz,Redder:1987xba,Ouellet:1992zz,Roters:1999zz,Gialanella:2001ayx,Kunz:2001zz,Assuncao:2006vy,Makii:2009zz,Plag:2012zz}}
\label{fig1}
%\end{center}
\end{figure}

\subsection{Fits with a channel radius of 5.43 fm}

Proposed OSGA experiments\cite{suleiman:2014aa,Ugalde:2012eh,DiGiovine:2015lda,Gai:2018sip,Balabanski:2017qth} are expected to have several orders of magnitude improvement in luminosity over previous experiments and should provide data at the lowest practical values of energy.  We take our best $R$-matrix fit of the $E1$ and $E2$ $S$-factor data as the most probable description of the projected JLab data.  We then randomly varied these OSGA $S$-factor pseudo data based on their projected uncertainties according to a Gaussian probability distribution about the best fit $S$-factor values.  The parameters that were used to provide the $R-$matrix curves shown in Fig.~\ref{fig1} are given for reference in Table~\ref{one}.
In order to study the impact of proposed OSGA data and low energy data in particular, we performed five fits:  a fit to existing $E1$ and $E2$ data (denoted by ``all" in Table~\ref{two}); a fit to data published after the year 2000 (denoted by ``2000"), both with (denoted by ``J" in table~\ref{two}) and without projected JLab data; and a fit to all data in Fig.~\ref{fig1} above 1.6 MeV (denoted by ``$E>$1.6" in table~\ref{two}).  Although it has been customary\cite{Perez:2016aol} to eliminate data sets that deviate by more than three standard deviations from the fitted results, we chose to select data sets after the year 2000 as a test of systematic deviations and as suggested by Strieder\cite{strieder}.  This approach assumes that experimental equipment and methods have improved over the decades.  Another reason for this approach is that not all authors of the data sets disclose their systematic errors.  The $S$ factors projected to 300 keV along with standard deviations, $\sigma$, which represent the statistical fit uncertainty are given in Table~\ref{two} for the five cases.  The reduced $\chi^2$ for the fit to the original data is also shown.   As a test of the method, we arbitrarily reduced the error bars for the projected JLab data by an order of magnitude and present the results as ``all J/10" in the table.   

 Several observations can be made from Table~\ref{two}.  The standard deviations for the total projected $S$-factors with proposed JLab data are generally smaller than those without JLab data.  The total and $E1$ projections appear to be significantly larger for $E>$1.6 MeV data than the fits to ``all" data, indicating the importance of low-energy data.  As expected the standard deviations for the ``all J/10" case are significantly smaller than that for the other cases.  For the fits to the data after 2000, the reduced $\chi^2$ is significantly smaller than that for fits to ``all" data.  This indicates that the data sets after 2000 are more consistent with one another than with all data sets.  Finally, the $S$-factor projections for $E2$ appear to be about a third of those for E1.  

As an example, the projections from the simultaneous fit to all  $E1$ and $E2$ data, the case represented by the first line in Table~\ref{two}, are shown in Fig.~\ref{fig2}. 
The dashed vertical line indicates the projection for the fit to the original data, while the histogram represents the results of fits to 1000 sets of randomized pseudo-data.  The dotted curve is a Gaussian based on the mean and standard deviation found from the fits.  The $S$(300 keV) from the fit to original data is 112.3 keV-b while the mean for the fits to pseudo-data is 114.0 keV-b.  The standard error for the fits to pseudo-data is about 0.2 keV-b.  Thus, the statistical error in the fits to 1000 sets of pseudo-data cannot alone explain the discrepancy.  If one speculates that the systematics in the original data are driving the discrepancy, then we could compare the ``2000" data.  The $S$(300 keV) for the fit to the ``2000" data is 123.5 keV-b, while the mean of the pseudo-data fits is 123.2 keV-b, in better agreement with one another.

\begin{figure}
\begin{center}
\includegraphics[width=3.4in]{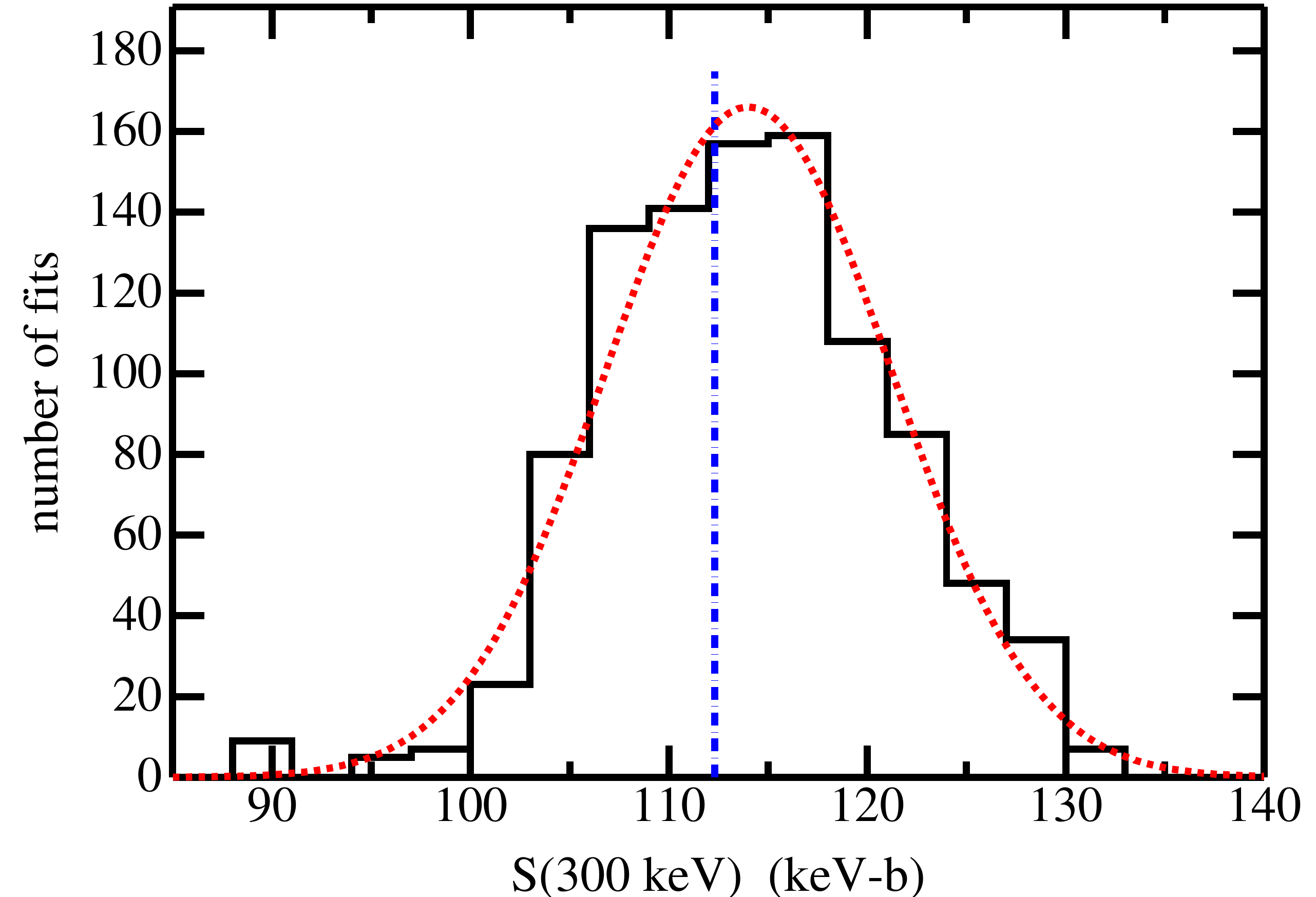}
\includegraphics[width=3.4in]{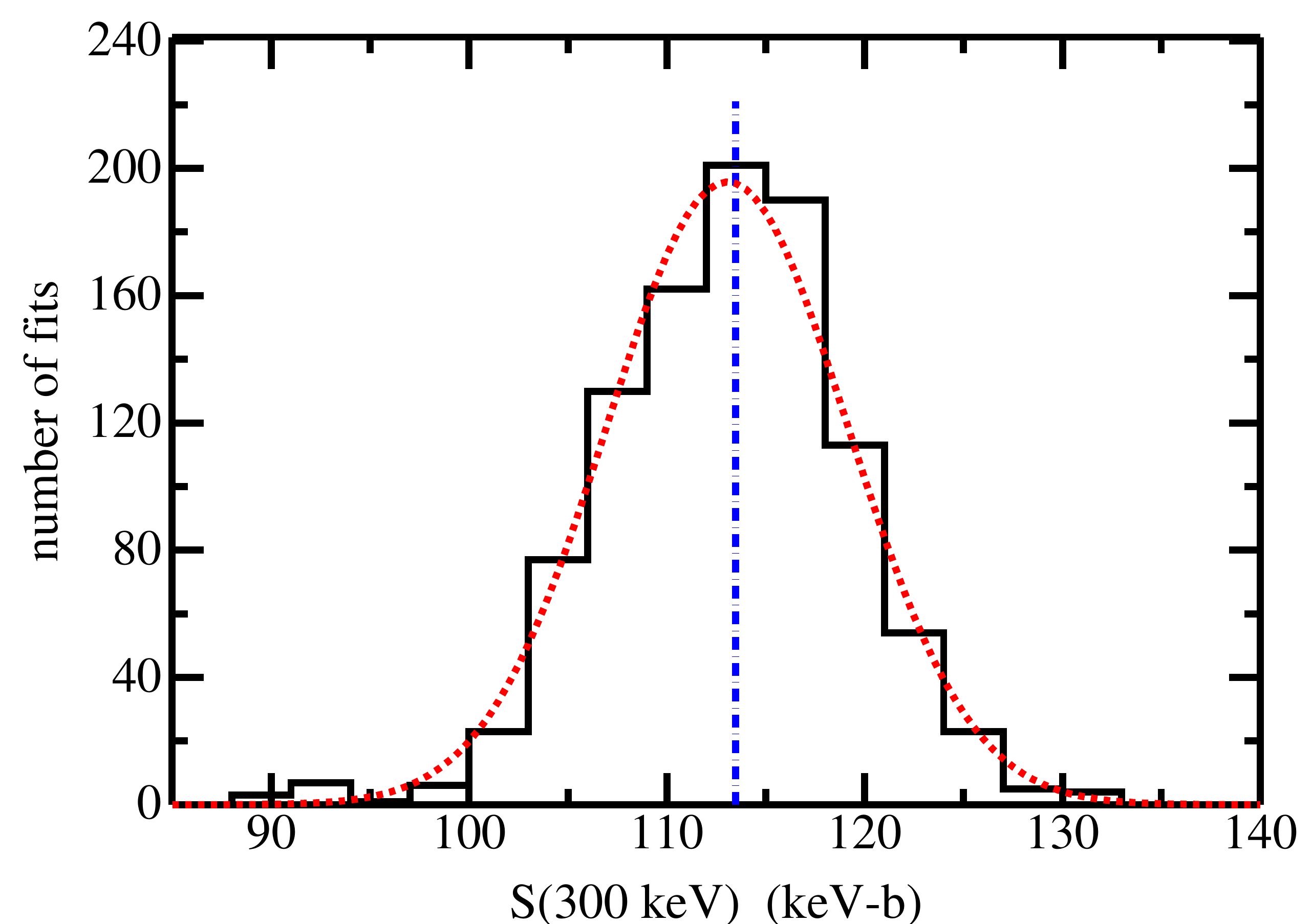}
\caption{Projections of the astrophysical $S$ factor to 300 keV for simultaneous fits of existing $E1$ and $E2$ data (top panel) and for $E1$, $E2$ and proposed JLab data (bottom) for a channel radius of 5.43 fm. The blue dashed vertical lines indicate the projections for the fit to the original data, while the histograms represent the results of 1000 fits to randomized data that would lie along the fit to original data.  The red dotted curves are Gaussians based on the means and standard deviations found from the fits. }
\label{fig2}
\end{center}
\end{figure}

%\begin{figure}
%\begin{center}
%\includegraphics[width=3.4in]{a_g11_two_543_09_30_2018.pdf}
%\caption{Histograms of the E1 sub-threshold reduced width amplitude for the ``all" data case, solid curve, and the ``E$>$1.6 MeV" data case, dashed curve.  The red dashed and blue dash-dotted vertical lines indicate the projections for the fit to the original data for the ``all" and ``E$>$1.6" MeV data, respectively. }
%\label{fig3}
%\end{center}
%\end{figure}

\begin{table}[!htp] %add [H] placement to break table across pages
\caption{\label{two}$S$-factor projections to 300 keV and standard deviations for total $S$, $S_{E1}$ and $S_{E2}$ for fits with a channel radius of 5.43 fm.}
\begin{ruledtabular}
\begin{tabular}{l c c c  c c c c}
data & orig $\chi_\nu^2$ & $S$  & $\sigma$  & $S_{E1}$  &   $\sigma_{E1}$  &  $S_{E2}$  &   $\sigma_{E2}$  \\
              &  &  &  &  &  (keV-b)    \\
\hline
all            &    2.3      & 112.3           &  7.2        &   77.6         &     6.4        &    34.7     &   2.8  \\
all J          &    2.2       & 113.5            &  6.1        &   81.8       &    5.8       &     31.7    &    3.0  \\
2000        &    1.7       &  123.5           &  6.9        &    89.6      &    6.4        &    33.9     &   3.3   \\
2000 J      &    1.7       &  125.0           &  6.7        &    89.7      &    6.3        &    35.2    &   3.4   \\
$E>$1.6   &    2.6       &  119.6            &  5.8       &   87.1       &    5.4        &    32.5    &   2.6   \\
\hline
all J/10    &     2.4      &  116.4   &         2.4            &  81.1        &    3.5      &      35.3    &   1.9   \\
all J/2 &    2.2      &  118.8            &    4.2     &    81.8       &   3.9         &    37.2    &   2.8    \\
\hline
\end{tabular}
\end{ruledtabular}
 \end{table}

\begin{figure}
\begin{center}
\includegraphics[width=3.4in]{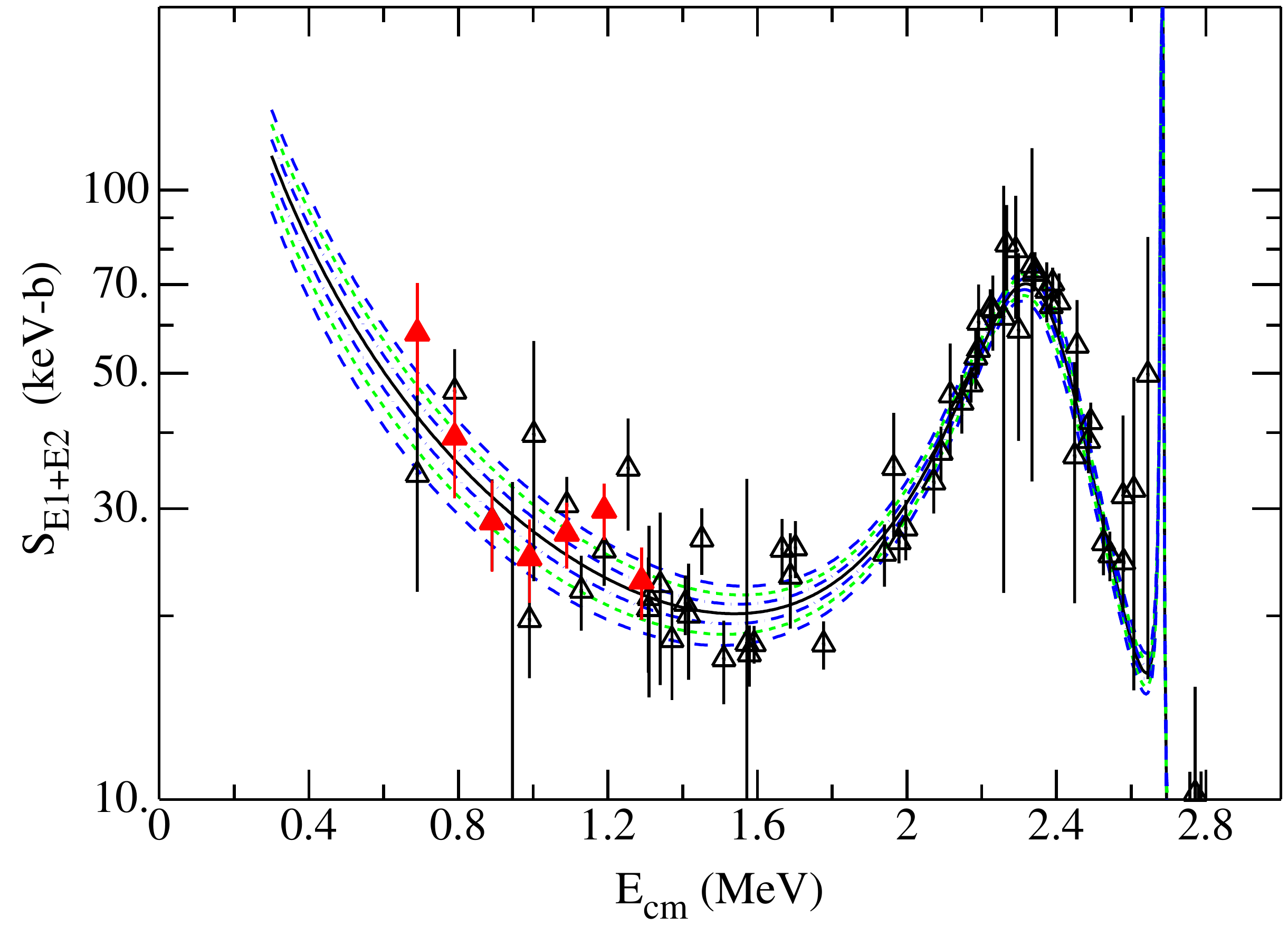}
\caption{Energy dependence of $S_{E1}$ + $S_{E2}$ from a fit to ``all" data indicating the $\pm$ 1, 2 and 3 standard-deviation bands shown as the dash-dot, short dash and long dash curves, respectively.  The curves are based on the parameters in Table~\ref{one}. The open triangles represent a sum of $E1$ and $E2$ where both E1 and E2 data exist.  The standard deviation at 300 keV is given by the first line and fourth column of Table~\ref{two}.  The projected JLab data are represented by the red triangles.}
\label{fig4}
\end{center}
\end{figure}

Fig. \ref{fig4} shows the curves that represent $\pm$ 1,2 and 3 standard deviation simultaneous fits to existing $E1$ and $E2$ data.  We generated the curves by performing 500 fits to the data, generating 500 sets of parameters similar to those in Table~\ref{one}, and then using the parameter sets to determine the standard deviation at each value of energy. The representative capture data, shown as open triangles, were taken as the sum of $E1$ and $E2$ results governed by where both $E1$ and $E2$ data exist.  The projected JLab data are represented by red triangles in the figure.  Given the statistical errors for the projected JLab data and the small number of values, one might not expect the projected JLab data to have a large impact on the statistical error.  Although the impact of new JLab data cannot easily be seen from this figure, reducing the expected JLab errors by only a factor of two could make a significant impact as illustrated by the last line in Table~\ref{two}.

\begin{figure}
\begin{center}
\includegraphics[width=3.4in]{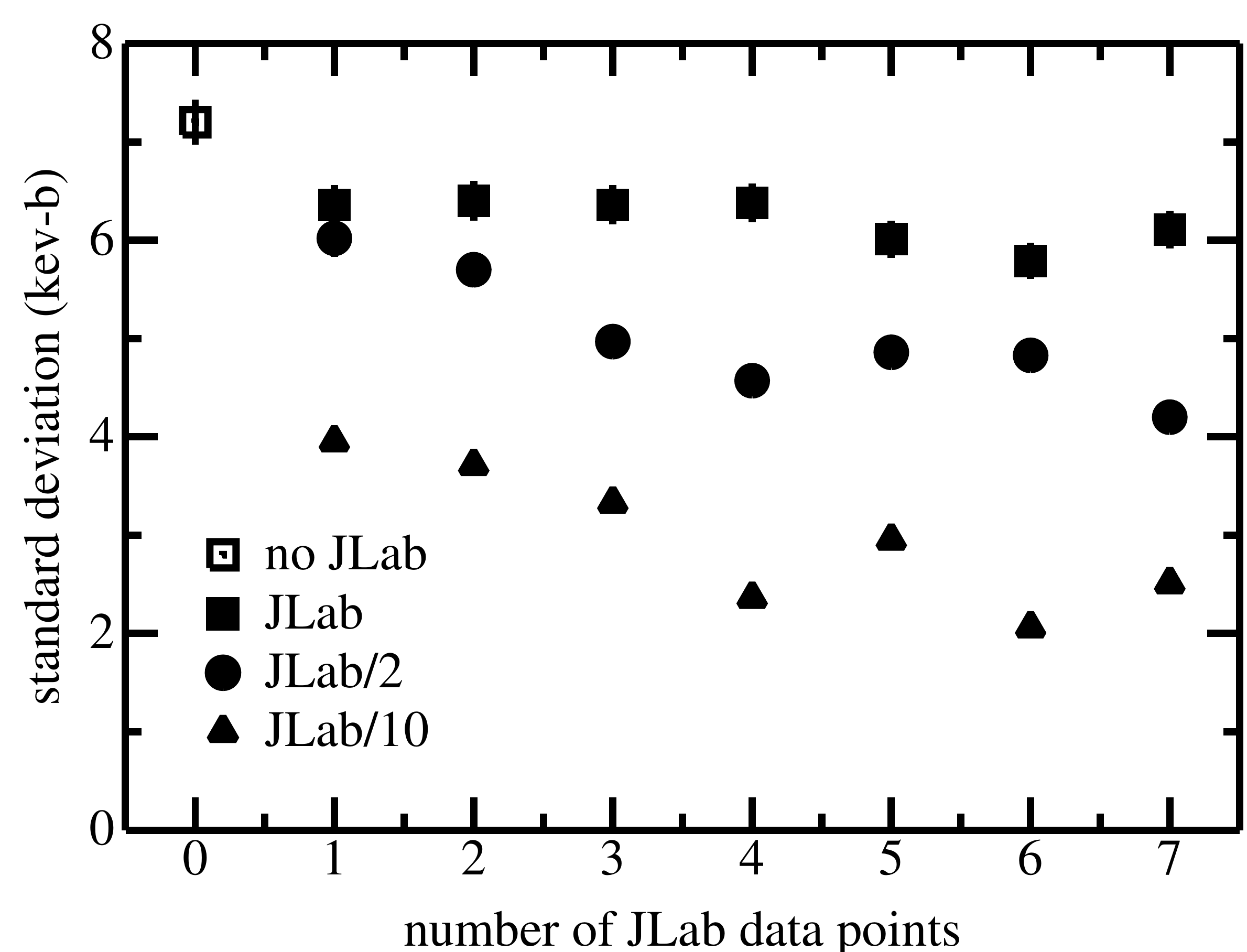}
\caption{Standard deviation from a 1000 fits with a channel radius of 5.43 fm to ``all" data with no JLab data (open square), with projected JLab data (solid squares) as a function of the cumulative number of JLab data points beginning with the highest energy JLab point, the same with JLab projected statistical errors divided by a factor of two (solid circles), and the projected JLab statistical errors divided by a factor of 10 (solid triangles).  The error limits shown in the figure are just the standard errors for the fits.}
\label{fig5}
\end{center}
\end{figure}

In order to more quantitatively explore the efficacy of the proposed JLab data, we made 1000 fits to a varying number of projected JLab data points from one to seven points beginning with the highest energy point 1190 keV and ending with the lowest energy point 590 keV.  These results are shown in Fig.~\ref{fig5}.  Note that we generated the JLab data as before by the fit values with a channel radius of 5.43 fm to ``all" data, then randomizing, according to the projected statistical errors.  We repeated this procedure with the JLab projected statistical errors divided by two as well as by ten.  These results are also shown in Fig.~\ref{fig5}.  The higher precision data indicate a clear pattern of diminishing returns in terms of the standard deviations as a function of the cumulative number of projected JLab data points.  This pattern is not so clear for the actual proposed JLab statistical errors. 

%\newpage
\subsection{Fits with a channel radius of 6.5 fm}

As mentioned before some previous $R$-matrix analyses have used a channel radius of 6.5 fm.  In order to be consistent with these previous analyses, we set the channel radius at 6.5 fm, and as before, we performed five fits:  a fit to existing $E1$ and $E2$ data (denoted by ``all" in table~\ref{four}); a fit to data published after the year 2000 (denoted by ``2000"), both with (denoted by ``J" in Table~\ref{four}) and without projected JLab data; and a fit to all data in Fig.~\ref{fig1} above 1.6 MeV (denoted by ``$E>$1.6" in Table~\ref{four}).  The $S$ factors projected to 300 keV along with standard deviations, $\sigma$, are given in Table~\ref{four} for the five cases.  The reduced $\chi^2$ for the fit to the original data is also shown.    As with the 5.43 fm case, the standard deviations for the total projected $S$-factors with proposed JLab data are generally smaller than those without JLab data.  Again, the total and $E1$ projections appear to be significantly larger for $E>$1.6 MeV data than the fits to ``all" data, and the size of the difference substantially exceeds the statistical errors.   As can be seen from comparing Tables~\ref{two} and \ref{four}, the $S$-factor projections to 300 keV are generally larger for a channel radius of 6.5 fm than those for 5.43 fm.  This finding is consistent with that of ref~\cite{deBoer:2017ldl}.  Again, the fit to data sets after 2000 also exhibit a smaller reduced $\chi^2$ than that for ``all" data.  It is interesting to note that if the errors on the expected 7 JLab data points are reduced by a factor of two, the case presented in the last line of Table~\ref{four}, then the result is in agreement with the 5.43 fm case, the first line in Table~\ref{two}.  This indicates that high quality data at low energy could even bring fits with different channel radii into agreement at least with regard to the extrapolation to 300~keV.

%The projections from the simultaneous fit to all  E1 and E2 data are shown in Fig.~\ref{fig5}. 
%The dashed vertical line indicates the projection for the fit to the original data, while the histogram represents the results of 500 fits to randomized data that would li%e along the fit to original data.  The dotted curve is a Gaussian based on the mean and standard deviation found from the fits.

\begin{table}[!htp] %add [H] placement to break table across pages
\caption{\label{four}Projections to 300 keV and standard deviations for total $S$, $S_{E1}$ and $S_{E2}$ for a channel radius of 6.5 fm.}
\begin{ruledtabular}
\begin{tabular}{l c c c  c c c c}
data & orig $\chi_\nu^2$ & $S$  & $\sigma$ & $S_{E1}$  & $\sigma_{E1}$  &  $S_{E2}$  &     $\sigma_{E2}$  \\
%\hline
                &               &   &  &  &  (keV-b)    \\
\hline  
all            &    2.3       &  124.9           &  8.3        &   80.6       &     7.1        &    44.3    &   5.0  \\
all J          &    2.2       &  121.7           &  6.3        &   84.3       &     5.9        &    37.4    &   2.8  \\
2000        &    1.6       &  131.3           &  8.3        &   90.5       &     7.7        &    40.7    &   3.8   \\
2000 J      &    1.6       &  131.3           &  7.2        &   90.5       &     6.9        &    40.7    &   3.8   \\
$E>$1.6 &    2.4      &   136.9           &  8.6        &   102.5       &    8.1         &   34.3     &   3.1    \\
\hline
all J/2   &    2.3       &  116.0           &  5.7        &   76.5       &     6.2        &    39.5    &   3.2   \\

\hline
\end{tabular}
\end{ruledtabular}
 \end{table}

\begin{figure}
\begin{center}
\includegraphics[width=3.4in]{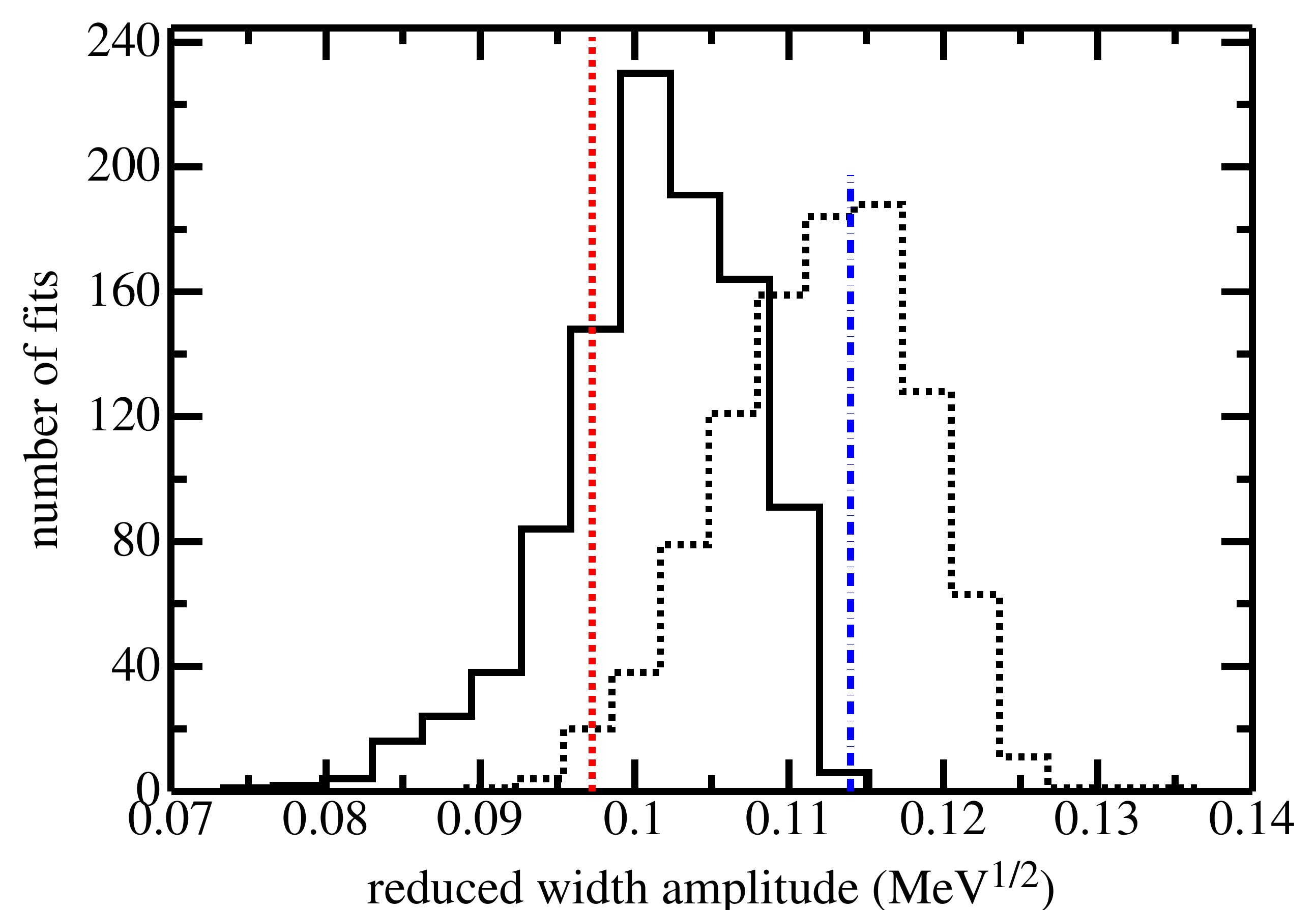}
\caption{Histograms of the sub-threshold E1 reduced width amplitudes the 6.5-fm fits for the ``all" data case, solid curve, and the ``E$>$1.6 MeV" data case, dashed curve.  The red dashed and blue dash-dotted vertical lines indicate the projections for the fit to the original data for the ``all" and ``E$>$1.6" MeV data, respectively. }
\label{fig6}
\end{center}
\end{figure}
The bound p-wave reduced width amplitudes found from the fits to ``all" and ``$E>$1.6" MeV data for a channel radius of 6.5 fm are given in Fig.~\ref{fig6}.  The histograms from the fits shown in the figure are asymmetric indicating that the error is not a Gaussian distribution.
The reduced width amplitudes of the bound $p$- and $d$-wave states, and the quantity $P_1\gamma_{11}^2$ found from the fits are given in Table~\ref{five} along with a recent value found from the 
$^{16}$N($\beta\alpha$) process\cite{Kirsebom:2018amt} for the bound $p$-wave state and for a transfer reaction\cite{Shen:2018eis} for the bound $d$-wave state.  Here, the quantity  $P_1\gamma_{11}^2$, where $P_1$ is the p-wave penetration factor evaluated at 300 keV, was included in table~\ref{five} in order to better compare with that of ref~\cite{Kirsebom:2018amt}. As pointed out in ref.~\cite{Kirsebom:2018amt} the quantity $P_1\gamma_{11}^2$ is the dominant term in the capture cross section.  The present fits  give values of  $P_1\gamma_{11}^2$ that are consistent with the experiment and analysis of ref.~\cite{Kirsebom:2018amt} although the channel radius of ref.~\cite{Kirsebom:2018amt}  is 6.35 fm.  The fits to data above 1.6 MeV (``$E>$1.6") give results that are larger for the $p$-wave state and smaller for the $d$-wave state than that for the other results.  Again, this indicates the importance of low energy data.  The fit for the after 2000 data that includes projected JLab data ``2000 J" reduces the statistical error somewhat for the bound $p$-wave state.  It is noted  that while the $S_{E2}$(300) of 46.2 $\pm$ 7.7 keV-b found from a recent transfer reaction\cite{Shen:2018eis} is in excellent agreement with the $S_{E2}$(300) from the present analysis with a 6.5 fm channel radius for ``all" data as indicated in Table~\ref{four}, the reduced width for the $E2$ bound state for the ``all" case differs by about two-sigma between these two approaches. 
\begin{table}[] %add [H] placement to break table across pages
\caption{\label{five} Reduced width amplitudes, $\gamma_{11}$  and  $\gamma_{21}$, and $P_1\gamma_{11}^2$ for the bound states from the fits to ``all", ``2000", and ``$E>$1.6" MeV data for a channel radius of 6.5 fm.  The result from $\beta$-delayed $\alpha$ decay of $^{16}$N  \cite{Kirsebom:2018amt} and for a transfer reaction\cite{Shen:2018eis} are also given for comparison.}
\begin{ruledtabular}
\begin{tabular}{l c  c c }
%   &        E1            &   E2   \\
Fit or data & $\gamma_{11}$   &  $P_1\gamma_{11}^2$ & $\gamma_{21}$ \\
    &  (MeV$^{1/2}$)  &  ($\mu$eV)  &   (MeV$^{1/2}$)   \\
\hline
all            &    0.097($^{+0.006}_{-0.005}$) & 4.68($^{+0.58}_{-0.48}$) &  0.150(9)\\
2000        &    0.104(($^{+0.006}_{-0.006}$) &5.38($^{+0.62}_{-0.62}$)   &  0.142(8)  \\
%2000 J      &    0.099(($^{+0.004}_{-0.004}$)     &    0.141(8)\\
$E>$1.6   &    0.114(($^{+0.003}_{-0.008}$)   &6.46($^{+0.34}_{-0.91}$)    &   0.130(6)\\
$^{16}$N($\beta\alpha$) &  -  & 5.17(75)(54)  &  - \\
$^{12}$C($^{11}$B,$^7$Li)$^{16}$O  &  -  & - &  0.134(18)  \\
\hline
\end{tabular}
\end{ruledtabular}
 \end{table}

\section{Summary}
 
From this study it appears that inverse reaction data can have a significant impact on the projection of $S$(300~keV) based on the projected OSGA data.
We took the projected JLab data to represent $E1$ + $E2$ data since only total cross sections to the ground state will be measured.  The projected standard deviation for the 1000 fits to the $E1$ and $E2$ data with the proposed JLab data is generally smaller than that without JLab data.   The JLab data constrain the total $E1$ + $E2$ cross section in the fit.  This leads to smaller standard deviations than fitting $E1$ and $E2$ separately.  Fitting only data above 1.6~MeV leads to a significant shift upward in the projected $S$-factors at 300~keV.  This illustrates the importance of lower energy data in the extrapolation to 300~keV.  Since the expected OSGA data will be less than 1.6~MeV and even lower than existing data, we can infer that the proposed OSGA data will have a significant impact on the value of the low energy extrapolation.  The significant difference between $S$(300 keV) for the fits with channel radii of 5.43 and 6.5~fm indicates model uncertainty.  The lower energy OSGA data may help resolve this ambiguity.   For example, if the uncertainties on the projected 7 JLab data points are reduced by a factor of two, the $S$(300 keV) from a fit with a 5.43 fm channel radius is brought into agreement with that from a 6.5 fm fit.  This level of accuracy at low energies would represent an interesting goal not only for the upcoming JLab experiment, but also for the other future experiments.

%\section{Acknowledgements}
\begin{acknowledgments}
We thank O. Kirsebom and F. Strieder for useful discussions.  This work is supported by the U.S. National Science Foundation under grant 1812340 and by the U.S. Department of Energy (DOE), Office of Science, Office of Nuclear Physics, under contract No. DE-AC02-06CH11357
\end{acknowledgments}

\bibliography{rmatrix_all_16O_1}

\end{document}